# In-fiber second-harmonic generation with embedded two-dimensional materials

Gia Quyet Ngo,[1*] Emad Najafidehaghani,[2] Ziyang Gan,[2] Sara Khazaee,[3] Antony George,[2] Erik P. Schartner,[4] Heike Ebendorff-Heidepriem,[4] Thomas Pertsch,[1,5,6] Alessandro Tuniz,[7] Markus A. Schmidt,[8] Ulf Peschel,[3] Andrey Turchanin,[2] and Falk Eilenberger[1,5,6*]

[1]*Institute of Applied Physics, Abbe Center of Photonics, Friedrich Schiller University Jena, Albert-Einstein-Str. 15, 07745 Jena, Germany*

[2]*Institute of Physical Chemistry, Abbe Center of Photonics, Friedrich Schiller University Jena, Lessingstraße 10, 07743 Jena, Germany*

[3]*Institute of Solid State Theory and Optics, Abbe Center of Photonics, Friedrich Schiller University Jena, Max-Wien-Platz 1, 07743 Jena, Germany*

[4]*ARC Centre of Excellence for Nanoscale BioPhotonics (CNBP), Institute for Photonics and Advanced Sensing, School of Physical Sciences, University of Adelaide, Adelaide SA 5005, Australia*

[5]*Fraunhofer-Institute for Applied Optics and Precision Engineering IOF, Albert-Einstein-Str. 7, 07745 Jena, Germany*

[6]*Max Planck School of Photonics, Germany*

[7]*Institute of Photonics and Optical Science (IPOS) and university of Sydney Nano Institute (Sydney Nano), School of Physics, The University of Sydney, Camperdown, NSW 2006, Australia*

[8]*Leibniz Institute of Photonic Technology (IPHT), Albert-Einstein-Str. 9, 07745 Jena, Germany*

*Corresponding authors:

Gia Quyet Ngo (quyet.ngo@uni-jena.de),

Falk Eilenberger (falk.eilenberger@uni-jena.de).

Silica-based optical fibers are a workhorse of nonlinear optics[1]. They have been used to demonstrate nonlinear phenomena such as solitons[2] and self-phase modulation[3]. Since the introduction of the photonic crystal fiber[4], they have found many exciting applications, such as supercontinuum white light sources[5-7] and third-harmonic generation[8,9], among others [10,11]. They stand out by their low loss, large interaction length and the ability to engineer its dispersive properties, which compensate for the small $\chi^{(3)}$ nonlinear coefficient. However, they have one fundamental limitation: due to the amorphous nature of silica, they do not exhibit second-order nonlinearity, except for minor contributions from surfaces. Here, we demonstrate significant second-harmonic generation in functionalized optical fibers with a monolayer of highly nonlinear $MoS_2$ deposited on the fiber's guiding core[12]. The demonstration is carried out in a 3.5 mm short piece of exposed core fiber[13], which was functionalized in a scalable process CVD-based process[14], without a manual transfer step. This approach is scalable and can be generalized to other transition metal dichalcogenides and other waveguide systems. We achieve an enhancement of more than 1000x over a reference sample of equal length. Our simple proof-of-principle demonstration does not rely on either phase matching to fundamental modes, or ordered growth of monolayer crystals, suggesting that pathways for further improvement are within reach. Our results do not just demonstrate a new path towards efficient in-fiber SHG-sources, instead, they establish a platform with a new route to $\chi^{(2)}$-based nonlinear fiber optics, optoelectronics and photonics platforms, integrated optical architectures, and active fiber networks.

While silica's small third-order nonlinearity is more than compensated by the fibers' low loss and the ability to engineer its dispersive properties, they do not exhibit bulk second-order nonlinearity, to its amorphous structure. The same applies to most non-silica fibers as drawing requires amorphous materials. Although second-order nonlinear responses can be induced, for instance, by electrical poling[15] using charged electrodes placed close to the core or by optical poling through irradiation with fundamental and second harmonic light[16,17], these approaches have not been found widespread use. A scalable functionalization of optical fibers with strong nonlinear optical materials and a large mode overlap, leading to a hybrid system with substantial second-order nonlinearities, would be a game-changer in terms of finally enabling much-needed efficiency in fiber-based harmonic sources, OPOs, and spontaneous parametric down-conversion sources.

In this work, we demonstrate a new, scalable, controllable, and reproducible approach to enable second-harmonic generation in fibers. We functionalize the surface of exposed core optical fibers (ECFs)[13] with a highly nonlinear monolayer of transition metal dichalcogenide (TMD)[18], similar to the process reported in[12]. Monolayer TMDs are a well-studied class of two-dimensional (2D) materials. Here we focus on the semiconducting monolayer $MoS_2$, because of its strong second-order optical nonlinearity[19,20]. First-principles analysis of the second-order nonlinear properties of monolayer TMDs predicts a comparable magnitude of $\chi^{(2)}$ to other highly nonlinear semiconducting bulk crystals[21], while experimentally reported $\chi^{(2)}$ values of $MoS_2$ vary significantly[19-25]. However, such values are all in the range of, or higher than, commonly used nonlinear bulk crystals. Moreover, second-harmonic generation (SHG) intensity is enhanced when the second harmonic (SH) wavelength overlaps with the C-exciton[19] or A and B-exciton band of 2D TMDs[26,27], all of which are in the visible spectral range and thus of immediate interest. For example, at resonance with the C-exciton, the sheet susceptibility $\chi^{(2)}_{MoS2,sheet}$ of the $MoS_2$ monolayer is estimated to be approximately $8 \times 10^4$ pm$^2$/V [19], which is extremely large when converting to the effective bulk second-order susceptibility of MoS₂: $\chi^{(2)}_{MoS2,eff} = \frac{c^{(2)}_{MoS2,sheet}}{t} =$ 123 pm/V [24], due to the layer's extremely small thickness of $t = 0.65$ nm.

The tiny thickness $t$, however, presents challenges when combined with the small interaction length on planar substrates, curtailing the overall nonlinear conversion efficiency and hampers applications. Therefore, and in line with a previous report on enhanced third-harmonic generation and in-fiber photoluminescence (PL)[12] in ECFs, we argue that the extension of the interaction length by the integration of MoS₂ in guided wave systems opens new application perspectives in nonlinear optics. Apart from nonlinear light conversion, this may extend into fiber-based sources for spontaneous parametric down-conversion[28] or remote sensing as SHG has been proven as an effective and non-destructive method to monitor the strain in 2D materials[29,30].

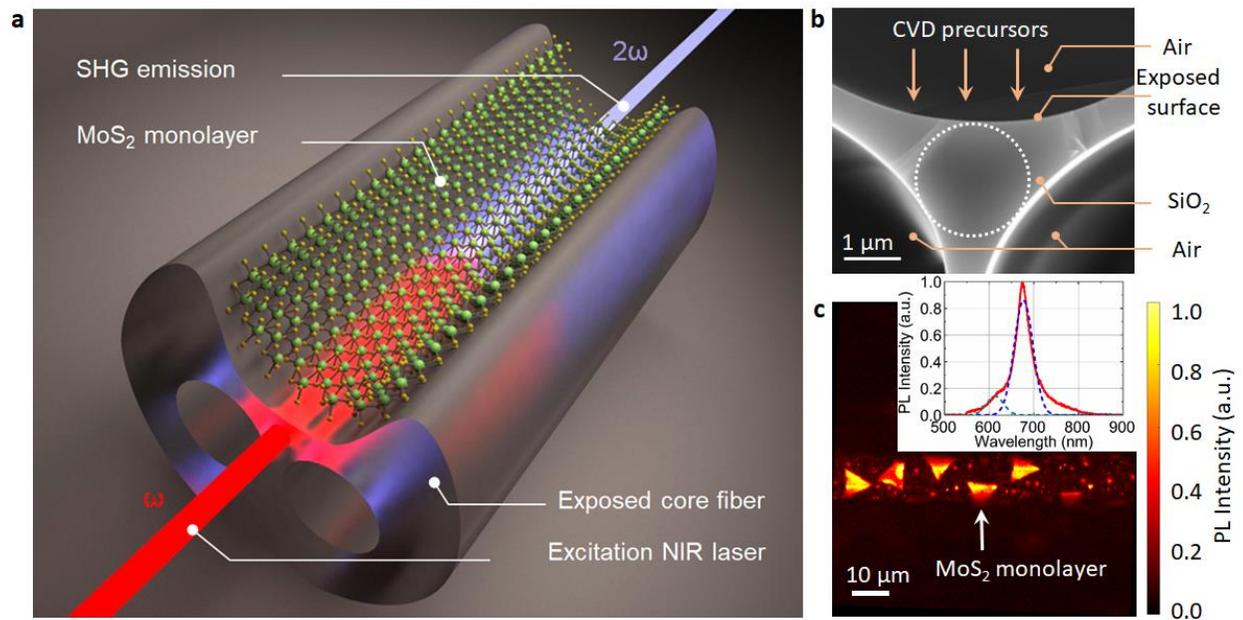

**Figure 1: Functionalization of ECFs with 2D materials. a,** Schematic of a SHG experiment with embedded 2D materials on an ECF. **b,** Cross sectional SEM image of the core region of the ECF. The dotted circle denotes the effective core area of the ECF. **c,** Top view PL mapping of a section of MoS₂-coated ECF. The inset shows PL emission spectrum from a typical monolayer of MoS₂ in **c** under the excitation of 532 nm laser after filtering by the long-pass filters at 550 nm.

Figure 1a shows a schematic summarizing the fundamental concepts of this work. The integration of highly nonlinear TMDs into the ECFs is carried out with a scalable and reproducible CVD process[14], where high-quality[31] TMD monolayer crystals are grown directly on the ECF's guiding core (radius: 1 μm). A cross-sectional scanning electron microscopy (SEM) image of an ECF is displayed in Figure S1a-b. Figure 1b shows a cross-sectional SEM image of the ECF's core. The precursors are carried by the gas flow inside the CVD reactor and start to grow on the exposed surface of the ECF's core. The core is supported by three thin struts of a homogeneous silica structure and surrounded by two air holes at the bottom and one open-air access on top of the functionalized surface. The crystal size and distribution are controlled by the precursor flow rate and the position of the ECF in the reaction zone. This results in densely distributed but randomly oriented monolayer crystals with an average size of 7 μm. Details on ECF fabrication are given in the Methods section. The small size of the guiding core leads to a small effective core area and increases the field overlap of the TMD crystals with the fundamental and the second harmonic modes. The overlap of the monolayer location with the evanescent field of the guided mode leads to a direct interaction of guided light and the TMD crystals. The fiber is hence functionalized and the optical properties of the TMD crystals can be used in a guided wave geometry.

Most previous works in the integration of monolayer TMDs with fibers and waveguides relied on the mechanical transfer of TMDs onto nanostructures[32] or waveguides[33], limiting scalability and reproducibility. In a similar approach, Zuo et al. have demonstrated CVD-based two-step growth of TMDs crystals in hollow-core and PCF fibers[34]. They demonstrate guided-wave SHG in the hollow core geometry. However, hollow core fibers must have a large modal cross-section to limit the propagation loss[35]. Moreover, the location of the functionalization layer is coincident with a zero of the guided wave. Both effects lead to a small field overlap between the fundamental wave and the TMD monolayer. The generated SHG will also contribute mostly to radiation modes. We, therefore, argue that our approach leads to much higher interaction efficiency and much lower power requirements for a given length of the fiber.

We stress that our work is a proof-of-principle demonstration of light conversion in 2D functionalized waveguides and many parameters remain to be optimized for possible future applications. The ECFs used here exhibit phase matching only to high-order modes. Also, the nonlinear crystals are randomly oriented and act as individual SH-sources, thus preventing a coherent build-up of the second-harmonic field. Moreover, no attempt to optimize the overlap between the modal field and the 2D materials has been carried out and resonant enhancement, e.g. operation at excitonic wavelengths, was not utilized due to the wavelength-fixed pulsed laser.

## RESULTS AND DISCUSSION

### Fiber Characterization

The outstanding advantage of 2D functionalized waveguides is that the linear properties including the dispersion and guided modes remain virtually unchanged, except for higher losses due to the absorption of 2D crystals, as can be seen from our calculation in Supporting Information Figure S3a. The enhancement of the nonlinearity in these waveguides is solely due to the action of the $MoS_2$ monolayer at the interface[12].

In the first step, we used the previously established method[12] to characterize the location, quality, and distribution of $MoS_2$ on the fiber and their interaction with the core mode. By performing atomic force microscope (AFM) imaging, we found a thickness of 0.9 nm (see Figure S2a-b in Supporting Information). The Raman spectrum of the $MoS_2$ crystals displays a characteristic spacing of 20.5 $cm^{-1}$ between two modes, which confirms the successful deposition of monolayers of TMD crystals on ECF's core region. We further carried out PL mapping along the ECF core, as displayed in Figure 1c. This was performed by a confocal PL lifetime microscope (Picoquant Microtime 200). Typical $MoS_2$ crystals with a size of roughly 7 μm can be seen in the image. The inset exhibits the PL spectrum of this crystal shown in Figure 1c. Spectroscopic analysis was used to unravel the relative contributions of the two exciton species observed in $MoS_2$. The PL emission reveals the exciton peaks at 677 nm (the dotted blue line) and 613 nm (the dotted cyan line) with a spectral full width at half maximum of 45 nm for both. These values are comparable to those of high-quality crystals grown on planar substrates by the same technique[14].

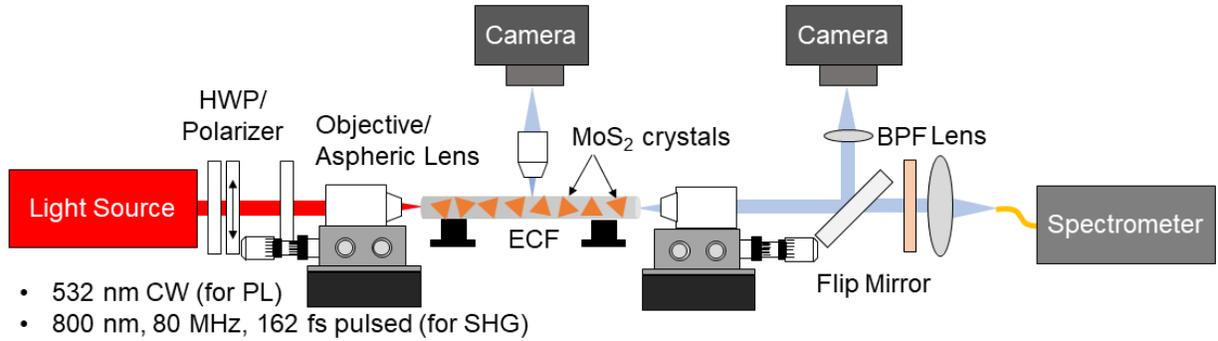

**Figure 2: Schematic diagram of experimental setup for PL, and SHG measurements.** HWP: half-wave plate, BPF: band-pass filter.

Because a large overlap between the guided modes and the monolayer crystals is crucial for efficient nonlinear interaction, we investigated the coupling of the TMDs to the core mode of the ECF. Figure 2 shows a schematic of the experimental setup used. The excitation light source was a 532 nm laser (Lighthouse Photonics Sprout) focused on the fiber core by a 40x microscope objective. The incident power was varied using neutral density filters, which are not shown in Figure 2. To achieve systematic statistics of the crystals ensemble on the fiber core throughout the fiber length, we image the PL emission perpendicular to the ECF using a microscope objective, a camera, and a pair of 550nm long-pass filters. Scanning the setup along the length of the fiber allows us to obtain a series of images, which we can stack to record the lengthwise distribution of PL-active monolayer crystals. Such a stack is depicted in Figure S5b. Tweaking the fabrication process, we obtained an almost 8x higher density of $MoS_2$ monolayers on the fiber core with the coverage up to 43.4%, as opposed to 5.4% in previously reported results[12].

## Second-harmonic generation

The generation of the second harmonic is governed by the interplay of second-order susceptibility, mode matching, and phase matching within the length of a flake of 7 μm. Because SHG is an intrinsic property of the TMDs, it also serves to quantify the material quality and layer thickness. Nonlinear experiments were carried out with a pulsed Ti:Sapphire laser (Coherent, 80 MHz repetition rate, 162 fs pulse width, see Figure S4 in Supporting Information), the output of which was coupled into the fiber core using an aspheric lens with the focal length of 3.1 mm. Light leaving the fiber was collimated with a 40x objective and coupled into a spectrometer (Horiba Jobin Yvon Triax) with a cooled Si-CCD-detector to measure the fundamental wavelength (FW) and SH spectra after passing through a series of band-pass filters at 400 nm. The power was controlled by a combination of a half-wave plate and a linear polarizer as illustrated in the schematic experimental setup. The input polarization was adjusted by a second half-wave plate in front of the aspheric lens.

While bulk silica itself has zero second-order susceptibility, there is a minor contribution stemming from the surface of the ECF core. This surface imposes a symmetry breaking perpendicular to the optical axis which in turn leads to SHG. We use the second-harmonic power generated by the bare ECF as a reference to compare against the $MoS_2$-coated ECFs. Both the power of the SHG and the driving FW were measured simultaneously. As previously discussed, the $MoS_2$-coated ECF has a higher loss per unit length than the bare ECF due to the absorption and scattering of TMDs. To minimize loss-related effects and the impact of the nonlinearly induced break-up of the pump pulse, we used a short fiber of 3.5 mm length. Figure 3a displays the recorded spectra of the FW for an averaged input power of 80 mW. The generated SH spectra are displayed in Figure 3b and demonstrate that SHG in the $MoS_2$-coated ECF is much higher than that of a bare ECF, irrespective of the input power. For ease of presentation, the SH intensity from bare ECF has been scaled up by a factor of 50. We also note that the SH spectrum of the $MoS_2$-coated fiber differs in shape from that of the plain fiber, most likely due to the frequency dependence of the nonlinear response of the $MoS_2$ crystals. Figure 3c shows the measured SHG in coated ECFs, revealing a quadratic dependence on the input power as expected.

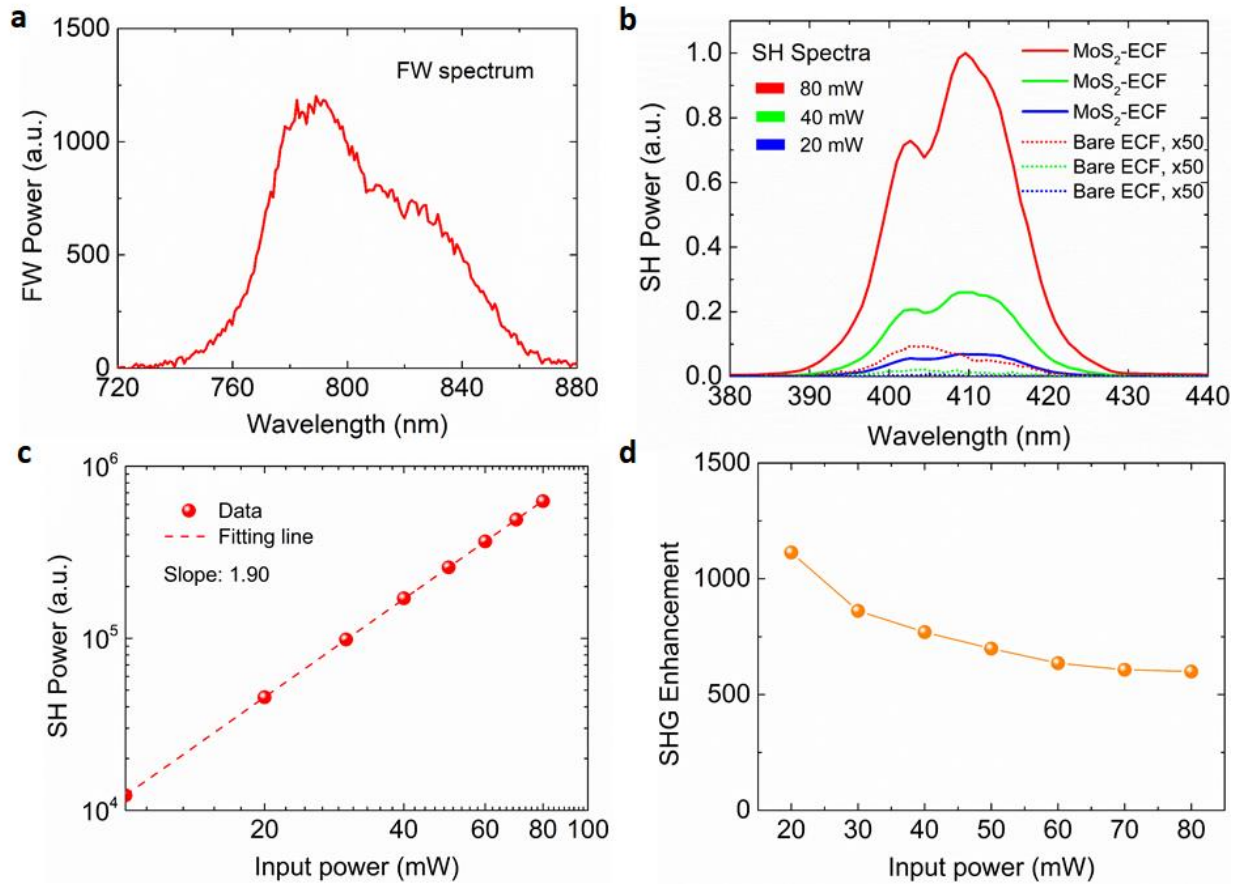

**Figure 3: SHG for horizontally polarized input light with respect to the functionalized surface in the center of the ECF. a**, FW spectrum of $MoS_2$-coated fiber at an input power of 80 mW. **b,** Normalized SHG spectra of both bare and coated fibers with three different input powers. The SH power from bare ECF was multiplied by a factor of 50 for ease of presentation. **c,** Log-log plot between SHG power vs input power for the $MoS_2$-coated fiber. **d,** Comparison of

SHG enhancement $\epsilon_{\text{SHG}} = P_{\text{MoS}_2}^{\text{SHG}}/P_{\text{Bare}}^{\text{SHG}}$ for a range of input powers with the incident horizontal polarization.

Using the bare ECFs as a reference, we systematically investigated the SHG enhancement of our MoS$_2$-coated fibers. The comparison was made for identical input powers between a pair of equal-length ECFs by calculating the ratio of SHG power $\epsilon_{\text{SHG}} = P_{\text{MoS}_2}^{\text{SHG}}/P_{\text{Bare}}^{\text{SHG}}$. We observed an 1113-fold increase in SHG conversion efficiency for an input power of 20 mW slightly dropping to about 600-fold for 80 mW, as depicted in Figure 3d. This drop of enhancement is most likely caused by the nonlinear effects that the TMD layer has on the FW, which enhances FW pulse break-up for higher power levels and thus quenches SHG. The more than 1000x increase in SHG is a testament to the highly nonlinear nature of MoS$_2$, given that the fraction of power flowing in the MoS$_2$ monolayer is calculated to be $\approx 3 \cdot 10^{-5}$ (see Figure S3b).

The role of absorption by the TMD is revealed in a cut-back experiment, in which we investigate the dependence of the SHG efficiency on the fiber length (see Figure 4). First, we measured the SHG spectrum of the 12 mm long fiber for horizontal input polarization, then the fiber was cut to a shorter length of 3.5 mm. Compared to the long fiber, the SHG efficiency increased 2.5 times for the short fiber, which we attribute to reduced propagation losses experienced by both the FW and SH waves, caused by the absorption in the MoS$_2$ crystals.

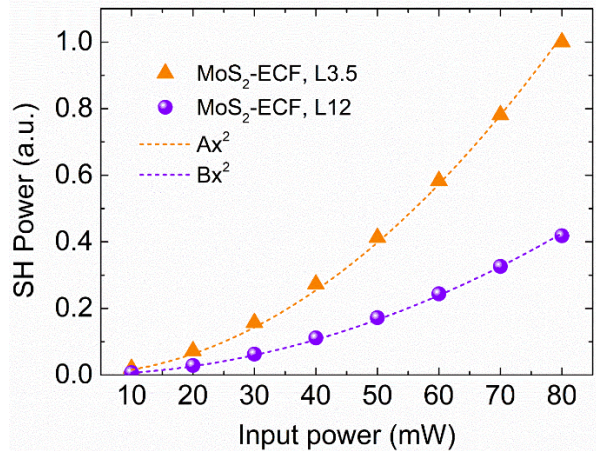

**Figure 4: Normalized SHG power of 3.5 mm long MoS$_2$-ECF vs 12 mm long MoS$_2$-ECF measured with a cut-back technique**. Two quadratic polynomial fits of the form $P_{SHG} = AP_{in}^2$ are shown ($\bar{R}^2 = 0.999$), where $P_{SHG}$ is the second harmonic generated power and $P_{in}$ is the input power from the femtosecond laser.

## SHG Modelling

To understand the process in more detail we computed the guided modes in the ECF core with a 0.65 nm thin MoS$_2$ deposited on the surface. The ECF supports multiple guided modes at both frequencies, while only the two fundamental modes (FMs) with horizontal and vertical polarization (see Figure 5a) have been excited.

Because phase matching plays a vital role in SHG, the effective mode indices of FMs and SH modes have been calculated within the spectral range of the applied pulses (grey area in Figure 4b). SHG occurs within the length scale of a single flake when the phase matching is satisfied, so that the relevant modes must approximately fulfill the condition $\Delta n = \left|n_{\text{fund}}^{\text{FW}} - n^{\text{SH}}\right| < \lambda_0/(\pi L_c)$, with $\lambda_0 = 800$ nm being the fundamental wavelength and $L_c \approx 7$ µm the mean length of a $MoS_2$ crystal (Figure S5a), yielding $\Delta n < 0.03$ (see the yellow band in Figure 5b). Here, we consider only the two fundamental modes for the FW, because only these are excited.

As displayed in Fig 4b, a significant number of modes fulfill this criterion, but the conversion efficiency is critically determined by field profiles and modal overlaps. All modes carried by the ECF are almost transversal and linearly polarized. In particular, the 2 µm-core of ECFs carries two virtually degenerate fundamental modes (FMs) of orthogonal polarization as shown in Figure 5a. Hence, the input polarization of the FW is maintained inside the ECF. Because TMDs only possess an in-plane nonlinear response, only the projection of the FW field onto the fiber surface is relevant, since this is where the TMD crystals are attached. Once generated, the SH polarization of the TMD will also oscillate along the surface of the ECF, but only its transverse component can excite a respective SH mode. Consequently, the total SH output is determined by a double projection, first of the FW onto the flake, and second of the SH field generated in the flake onto a respective SH mode, both being almost completely transversal. As the TMD flakes have random orientations, they on average contribute to SHG with the same efficiency but do not further influence the polarization dependence of the conversion.

The second harmonic power $P_{SH}$ emitted in a certain SH mode of the coated ECF is therefore approximately proportional to[36]

$$P_{SH} \propto \left| \int_{\text{surface}} ds \left[\vec{e}(s)\vec{E}_{FW}(s)\right]^2 \left[\vec{e}(s)\vec{E}_{SH}(s)\right] \right|^2 \tag{1}$$

with $s$ parametrizing the TMD coated surface of the ECF cross-section and $\vec{e}(s)$ being a transverse unit vector pointing along this surface. Equation (1) produces the required double projection of the FW $\vec{E}_{FW}(s)$ and SH mode fields $\vec{E}_{SH}(s)$. The specific values are calculated at $\lambda_0$ for all second harmonic modes, which have an effective index, that fulfills the phase matching condition. The results are displayed in Figure 5c.

The field structure of the FW displayed in Figure 5a together with the fact that TMD crystals are situated at the upper interface causes SHG to be more efficient for horizontal (*x*) polarization of the FW. This can be seen by comparing the blue and red bars in Figure 5c, which shows a stronger normalized overlap factor for the horizontal (*x*) polarization. However, as the upper surface of the ECF is strongly curved excitation with a vertically polarized input field remains non-negligible. The average ratio of the conversion efficiency can be determined by the ratio of the average height of the blue bar vs. the average height of the red bars, yielding $\varrho_M = 2.0$.

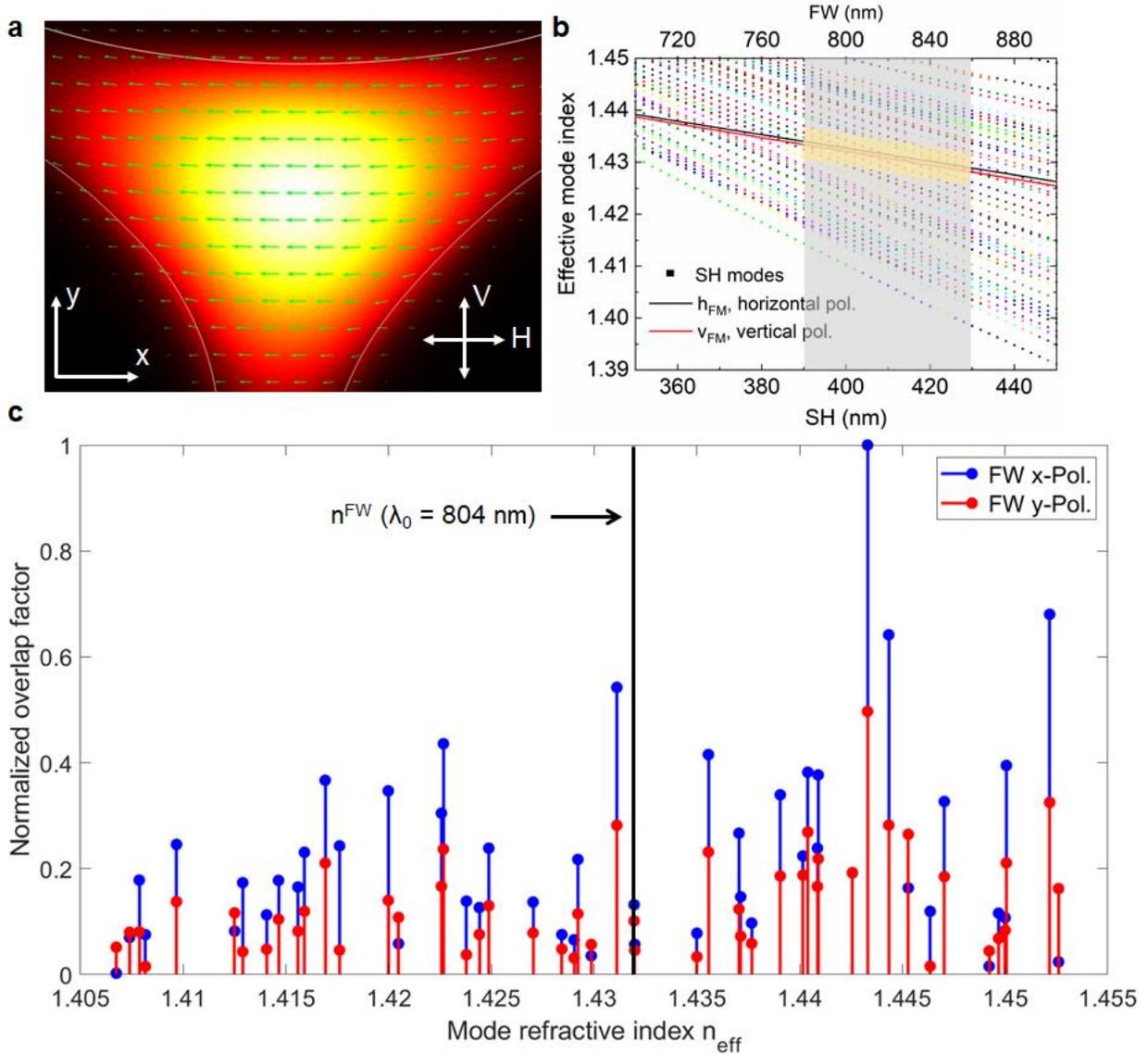

**Figure 5**: **Numerical simulation of SHG process in $MoS_2$-coated ECF. a**, Simulated transverse electric field norm of the horizontally polarized FM at 804 nm (The white double-headed arrows denote vertical and horizontal polarization). **b**, Effective mode indices of FMs (full lines) and SH modes (dotted lines) in the frequency between 700 - 900 nm and 350 - 450 nm, respectively (grey area: pulse spectrum, yellow area: phase-matching region with $\Delta n \leq 0.03$). **c**, Calculated normalized overlap factors for SH modes for both horizontal- and vertical-polarization of FW within the phase matching region. The black line is the phase matching point $n^{FW}$ = 1.432.

## Tunability of SHG by the input polarization

The above results indicate that SHG can be tuned by rotating the input polarization of the FW from horizontal (0°, in-plane with the functionalized surface) to vertical (90°, out-of-plane with the functionalized surface) as displayed in Figure 6a (black line). The power was fixed in this measurement and an HWP was used to change the input polarization of the excitation laser. As expected, the SH intensity is the highest for the horizontal polarization of the input beam. By altering the input polarization, the SH intensity drops to a minimum at 40° before increasing again at 90°.

This functional dependence is a result of the projection of the FH field onto the horizontal parts of the ECF surface (see equation (1)) scaling like the cosine, which is squared in SHG resulting in a total SH power proportional to the fourth power of the cosine of the input polarization angle. The same applies to vertical parts of the surface, where the cosine is replaced by the sine. Mixed angular terms (e.g. products of squared sine and cosine) depend critically on the geometry and are assumed to be almost negligible from the actual multimode ECF structure. Decomposition into these two components yields the expression:

$$P_{SH} \sim |\tilde{\chi}^{(2)}|^2 (\varrho \cos^4(\theta) + \sin^4(\theta)) \qquad (2)$$

More accurate calculations would require a better knowledge of the concrete structure of the ECF and a detailed understanding of the orientation and location of all individual crystals. Despite the simplicity of this model, the experimental results are well-approximated by this formula, as can be seen by fitting experimental data to equation (2) as shown in Figure 6a. Here, $\theta$ is the polarization angle with respect to the x-axis, $\tilde{\chi}^{(2)}$ is effective nonlinear coefficients resulting from the joint interaction of all possible SH modes, and $\varrho$ is the relative impact of the SHG excited by horizontal FW as opposed to the vertical FW. Here $\varrho = 1.23$ is used as a fitting parameter. A comparison with the value of $\varrho_M = 2.0$, determined from the calculated modes of the structure, is remarkably close to the measured value given the simplicity of the model. Our simple model is applicable for other curly functionalized surfaces of different waveguide systems and enables us to model the polarization of harmonic sources in similar optical fibers.

The complex multimode nature of the process is also underlined by the structure of the SH modes, which are mapped as a function of the input polarization and displayed in Figures 6 b, c, and d. It can be seen that neither of the intensity distributions are completely symmetric due to the asymmetry structure of the fibers themselves. It can also be seen that the SH field profile does not vary dramatically with input polarization. This is consistent with the data collected from Figure 5c, where the overlap coefficients can be seen to differ quantitatively between two SH modes but also that the ratio between horizontal (x) and vertical (y) input polarization of the same mode is roughly maintained across most of the modes.

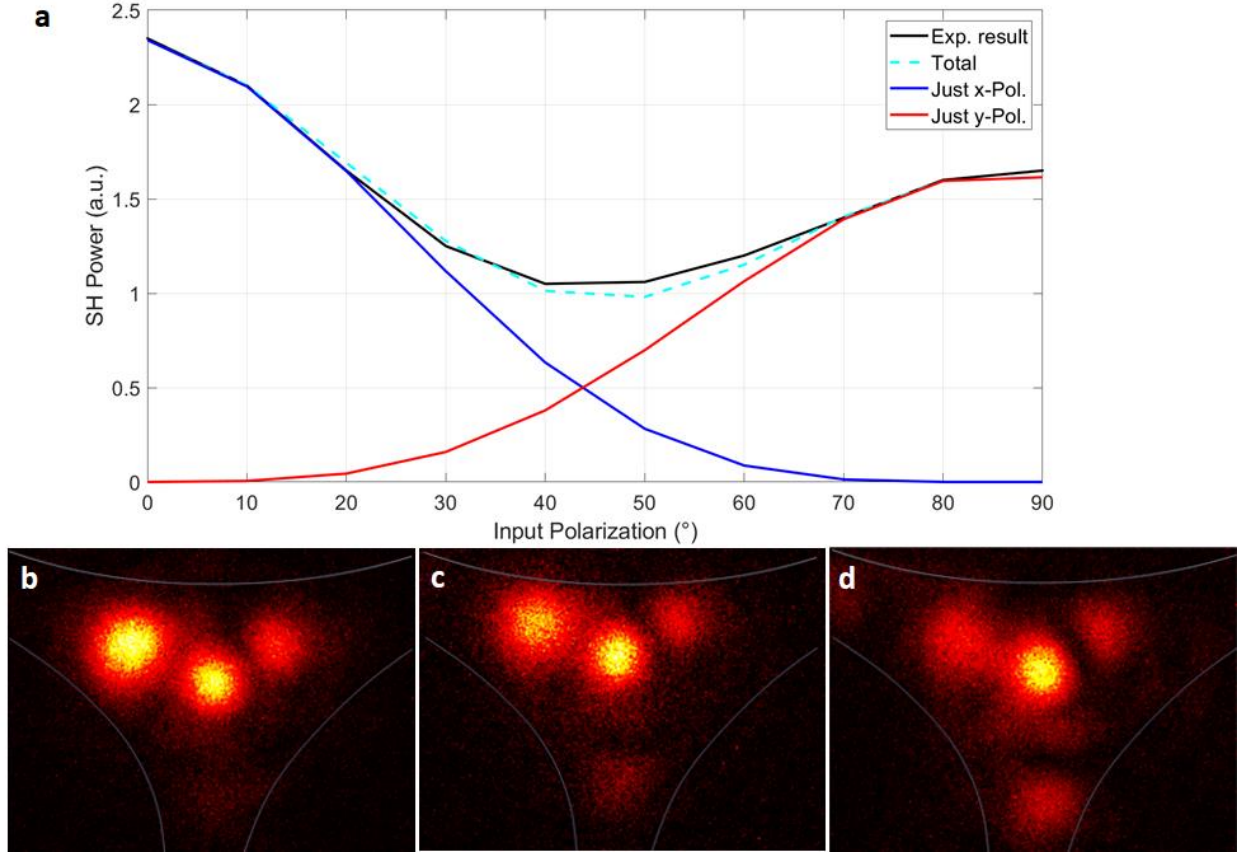

**Figure 6: SHG tunability. a**, SH intensity vs. input polarization compared with the functionalized surface, $0^0$ indicates horizontal polarization and $90^0$ denotes vertical polarization both experimental result and numerical calculation. x- and y-polarizations mean the near-degenerate fundamental modes with mostly horizontally or vertically polarization directions, respectively. **b-d,** Experimentally acquired images of the light distribution of SHG field at the facet of a $MoS_2$-coated ECF. **b**, Input polarization at $0^0$. **c**, Input polarization at $40^0$. **d**, Input polarization at $90^0$.

## CONCLUSIONS

In summary, we have demonstrated in-fiber second-harmonic generation from an optical fiber with a silica core by functionalization of the core with a highly nonlinear monolayer of CVD-grown $MoS_2$. While the demonstration was carried out in a non-optimized system we still observe a massive enhancement of the SHG by more than a thousand-fold. Further orders of magnitude enhancement can be expected by tuning of the phase matching, the optimization of the field overlap, and the growth of mutually oriented crystals or large monocrystalline films. The scalable CVD-based deposition process can be expanded to other fiber types and materials. It establishes a highly versatile photonic platform to further investigate the optoelectronic properties of 2D TMDs. Efficient, fiber-based, three-wave mixing components may now be possible, with application scenarios as SHG-sources, OPOs/OPAs, in optical signal processing, or for SPDC based photon-pair sources. The longer interaction length and unusual polarization geometry may also help explore excitonic properties, for instance, related to dark excitons in $XSe_2$-type materials. As SHG is a reliable indicator of strain, this work may also expand the applicability of fiber-based sensors and active fiber networks.

## METHODS

**ECF fabrication.** The all silica ($SiO_2$) ECF was manufactured using an ultrasonic drilled silica preform with the opening on one side of the central section of the fiber to make the exposed core. During the drawing process, the glass and air holes are heated to the glass transition temperature in a furnace at a typical temperature range of 1900° C - 2000° C. The preform was caned and inserted into a jacket tube, then drawn into an ECF. The optical fiber has the comparable geometry to the preform except for the smaller size and the much longer length. The fabricated fiber has an outer diameter of 175 μm and an effective core diameter of 2 μm. Because of the chemically inert and temperature stable process, silica fiber has minimal internal stress due to heating and cooling. It can sustain both the temperature and the atmosphere condition of the CVD process. More details about the ECF fabrication can be found in reference[13]. The cross-section and geometry of our ECFs are shown in Figure S1.

**CVD growth of $MoS_2$ on fibers.** $MoS_2$ crystals were grown on the ECFs by a modified CVD technique, where a Knudsen-type effusion cell was used to deliver the sulfur precursor. Metal oxide powder was used as the source of transition metal atoms. Details of the method are given in reference[14] with the usage of ECFs as the substrate. The grown $MoS_2$ on the ECFs were characterized using optical microscopy (Zeiss Axio Imager Z1.m) and Raman spectroscopy (Bruker Senterra spectrometer operated in backscattering mode using 532 nm wavelength obtained with a frequency-doubled Nd:YAG Laser, together with a 100x objective and a thermoelectrically cooled CCD detector).

**PL mapping and spectroscopy.** PL and crystal size were mapped with a commercial confocal PL lifetime microscope (Picoquant Microtime 200), using a 532 nm laser for excitation. The maps were achieved by moving the focus of the 100x microscope objective along the ECFs core. The device has a spatial resolution in the order of 0.5 μm. The fiber was placed flat on the specimen table and thus perpendicular to the optical path. The PL signal was detected by a single-photon avalanche diode after passing through series of band-pass filters. The PL spectra were recorded by a grating spectrometer (Horiba Jobin Yvon Triax) equipped with a cooled CCD camera.

**Transverse in-fiber PL mapping.** PL was excited with a 532 nm CW laser (Lighthouse Photonics Sprout). An sCMOS camera (Zyla 4.2 plus) was mounted laterally together with a 10x objective to image the sideways PL emission from grown crystals. A set of 550nm long-pass filters was utilized to reject the scattered excitation laser. The camera and the objective were mounted on a translational stage to map the entire length of the fiber.

**SHG Measurements.** Nonlinear experiments were performed with a femtosecond pulsed laser with a duration of 162 fs at a central wavelength of 800 nm at a repetition rate of 80 MHz (Ti:Sapphire pulsed laser Coherent) and focused into the fibers with an aspheric lens. The laser was strongly chirped (shortwave light arriving before longwave light), the transform-limited pulse duration is in the range of 20 fs. No attempt of chirp compensation was undertaken. The signal leaving the ECF was collimated with a microscope objective and coupled into a Horiba spectrometer (Horiba Jobin Yvon Triax) with a cooled Si-CCD detector to measure the FW and the SH spectra. A band-pass filter was utilized to cut off the laser excitation.

## ASSOCIATED CONTENT

**Supporting Information**

The Supporting Information is available free of charge at

## AUTHOR INFORMATION

**Corresponding Authors**

**Gia Quyet Ngo** (quyet.ngo@uni-jena.de)

*Institute of Applied Physics, Abbe Center of Photonics, Friedrich Schiller University Jena, 07745 Jena, Germany*

**Falk Eilenberger** (falk.eilenberger@uni-jena.de)

*Institute of Applied Physics, Abbe Center of Photonics, Friedrich Schiller University Jena, 07745 Jena, Germany*

**Author contributions**

MAS, ATur, TP, UP, and FE developed the concept. GQN conducted experiments and simulations. EN and ZG were responsible for material growth and structural characterization. SK, ATun, and UP supported the modeling of the nonlinear behaviors in exposed-core fiber. AG developed the CVD process and adapted it for the ECFs. ES and HEH developed the ECFs fabrication technique, provided samples, and supported the modeling of mode properties. All authors contributed to the manuscript.

**Notes**

The authors declare no competing financial interest.

## ACKNOWLEDGEMENTS

QGN and EN are supported by the European Union, the European Social Funds, and the Federal State of Thuringia as FGR 0088 under Grant ID 2018FGR00088. AG and FE are supported by the German Research Council as part of the CRC SFB 1375 NOA projects B2 and B3 respectively. FE was supported by the Federal Ministry of Education and Science of Germany under Grant ID 13XP5053A. ES and HE are supported by the ARC Centre of Excellence for Nanoscale Biophotonics (CE140100003). Fiber fabrication was performed in part at the OptoFab node of the Australian National Fabrication Facility utilizing Commonwealth and SA State Government funding. AT is the recipient of an Australian Research Council Discovery Early Career Researcher Award (DE200101041).

The authors thank Stephanie Höppener and Ulrich S. Schubert for enabling their Raman Spectroscopy studies at the JCSM and Tilman Lühder for supporting the cut-back technique.**REFERENCES**

# Supporting Information

# for

# In-fiber second-harmonic generation with embedded two-dimensional materials

Gia Quyet Ngo,[1*] Emad Najafidehaghani,[2] Ziyang Gan,[2] Sara Khazaee,[3] Antony George,[2] Erik P. Schartner,[4] Heike Ebendorff-Heidepriem,[4] Thomas Pertsch,[1,5,6] Alessandro Tuniz,[7] Markus A. Schmidt,[8] Ulf Peschel,[3] Andrey Turchanin,[2] and Falk Eilenberger[1,5,6*]

[1] Institute of Applied Physics, Abbe Center of Photonics, Friedrich Schiller University Jena, Albert-Einstein-Str. 15, 07745 Jena, Germany

[2] Institute of Physical Chemistry, Abbe Center of Photonics, Friedrich Schiller University Jena, Lessingstraße 10, 07743 Jena, Germany

[3] Institute of Solid State Theory and Optics, Abbe Center of Photonics, Friedrich Schiller University Jena, Max-Wien-Platz 1, 07743 Jena, Germany

[4] ARC Centre of Excellence for Nanoscale BioPhotonics (CNBP), Institute for Photonics and Advanced Sensing, School of Physical Sciences, University of Adelaide, Adelaide SA 5005, Australia

[5] Fraunhofer-Institute for Applied Optics and Precision Engineering IOF, Albert-Einstein-Str. 7, 07745 Jena, Germany

[6] Max Planck School of Photonics, Germany

[7] Institute of Photonics and Optical Science (IPOS) and university of Sydney Nano Institute (Sydney Nano), School of Physics, The University of Sydney, Camperdown, NSW 2006, Australia

[8] Leibniz Institute of Photonic Technology (IPHT), Albert-Einstein-Str. 9, 07745 Jena, Germany

*Corresponding authors:

Gia Quyet Ngo (quyet.ngo@uni-jena.de),

Falk Eilenberger (falk.eilenberger@uni-jena.de).

**Fiber Geometry**

ECFs belong to a special subclass of suspended core fibers and are classified as microstructured optical fibers due to the opening on one side of the central section of the fiber during the fabrication process. The entire fiber is made of silica and the core is supported by three thin glass struts, as can be seen in Figure S1a-b. The upper part of the core is exposed to the external environment and forming a trench that runs along the length of the fibers. SEM images of the entire fiber cross-section and zoom into the guiding core region are displayed in Figure S1a and S1b, respectively. The core is surrounded by the air holes at the bottom and the open air at the top to define the functionalized waveguide. The significantly long suspension of the three struts to the body of the fiber makes the tight confinement of the guided mode inside the effective core area.

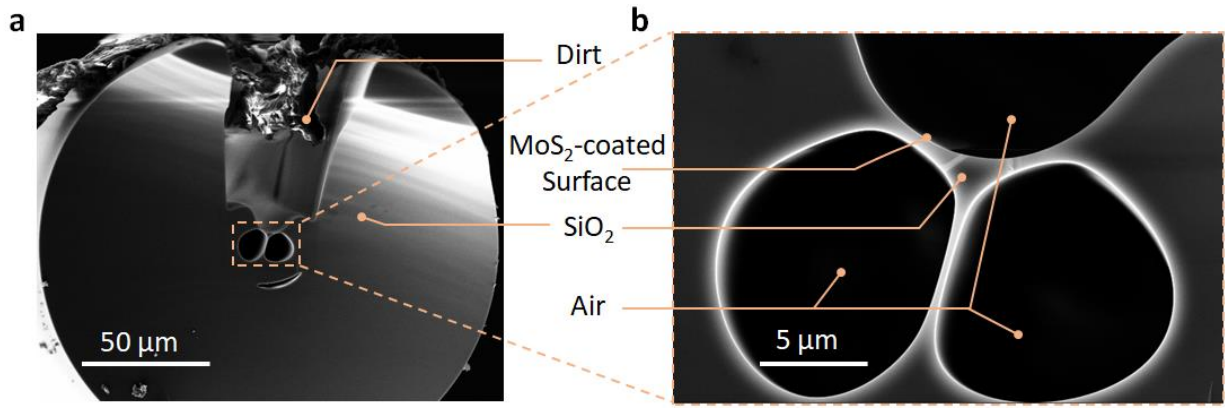

**Figure S1. a**, SEM cross section of an ECF. The exposed core and the groove running along the fiber length are marked in the orange rectangle. **b**, The enlarged cross section of the core and air holes of the ECF. The core ($SiO_2$) and the cladding (air) is visible at the orange box.

**Atomic force microscope (AFM) imaging and Raman characterization**

We grow $MoS_2$ crystals on the all-silica fibers by a modified CVD process where a Knudsen-type effusion cell is employed to control and deliver sulfur precursors. The morphological structure of a $MoS_2$-flake grown on the exposed side of ECF was mapped with an AFM over a range of roughly 10x10 µm. The AFM measurements were performed with a Ntegra (NT-MDT) system in tapping mode at ambient conditions using n-doped silicon cantilevers (NSG01, NTMDT) with resonant frequencies of 87 – 230 kHz and a typical tip radius of < 6 nm. The topological image is shown in Figure S2a with the typical trigonal geometry. The contaminations in the image were attributed to the long-term use in the experiment before AFM measurement and not related to the growth process. Figure S2b exhibits a height profile of the flake along the line marked in Figure 1a. The measured height of the flake was 0.9 nm, which is a typical thickness of a $MoS_2$ monolayer. The grown $MoS_2$ crystals were further characterized with Raman spectroscopy as depicted in Figure S2c (Bruker Senterra spectrometer operated in backscattering mode using 532 nm wavelength obtained with a frequency-doubled Nd:YAG Laser, a 100x objective, and a thermoelectrically cooled CCD detector). Figure S2c shows a typical Raman spectrum of the $MoS_2$ crystals grown on the exposed side of ECF with a characteristic spacing of 20.5 cm$^{-1}$ between the Raman resonances. This spacing feature is consistent with other works for CVD-grown monolayer TMDs on the planar substrate[1,2].

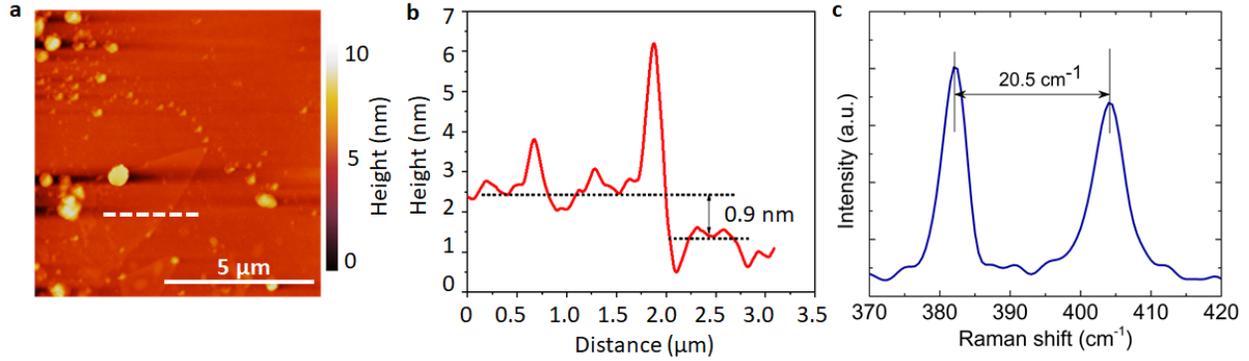

**Figure S2. AFM image and Raman spectrum of a typical CVD-grown MoS$_2$ monolayer on the SiO$_2$ wafer. a**, AFM topography map. **b**, AFM height profile as measured along the dashed line in **a**. **c**, Raman spectrum of the MoS$_2$ crystal displayed in **a**.

### Properties of fundamental guided modes

The numerical calculation of bare ECF and MoS$_2$-coated ECF has been discussed in our previous work[3]. Group velocity dispersion and attenuation properties of the coated ECF were calculated using finite element simulations (COMSOL v5.5) in Figure S3a. The ECF core is made from silica and therefore is set as a lossless medium in our calculation. The loss of the MoS$_2$-coated ECF is produced by the high extinction coefficient of the MoS$_2$ layer. Propagation loss was experimentally measured. For the MoS$_2$-coated ECF, the transmitted power is roughly 50% lower than the bare ECF at the identical setting. A cut-back measurement of a typical MoS$_2$-coated ECF with a length of 33 mm yielded a net loss of 0.12 dB/mm. Scattering boundary conditions were applied to analyze FM of fundamental wavelengths and high-order modes of second-harmonic wavelengths. Power flow energy in MoS$_2$ monolayer, for instance, the z-component of the Poynting vector of the fundamental mode is displayed in Figure S3b.

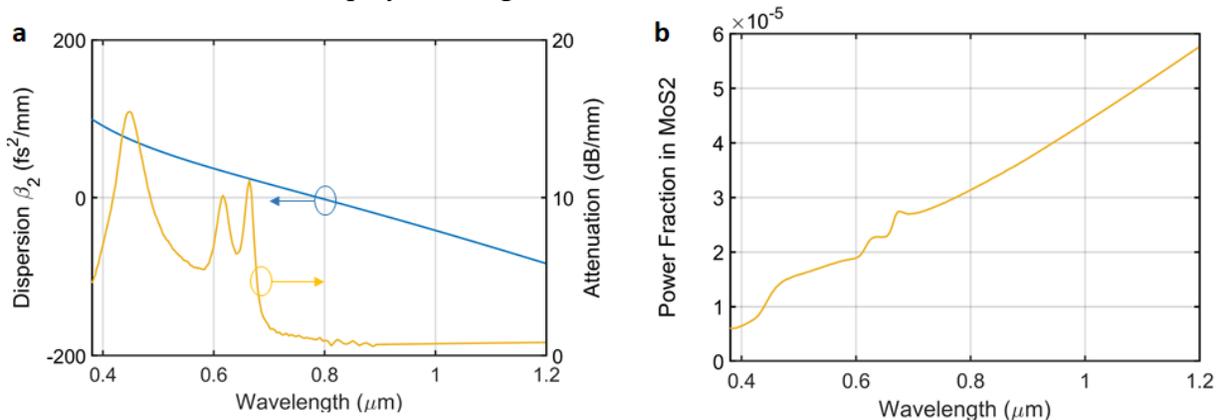

**Figure S3. a**, Dispersion (blue) and attenuation (orange) of the MoS$_2$-coated ECF. **b**, Fraction of the modal power flow (for instance fraction of the integral over the z-component of the Poynting vector), occurring in the single layer of MoS$_2$.

### Laser Pulse Characteristic

The pulse duration was estimated by measuring the spectrum and the non-collinear intensity autocorrelation (AC) of the pulse incident to the fiber (see Figure S4). The final estimate for the pulse duration of 162 fs at the fiber facet was obtained by further numerical propagation through the used aspheric lens.

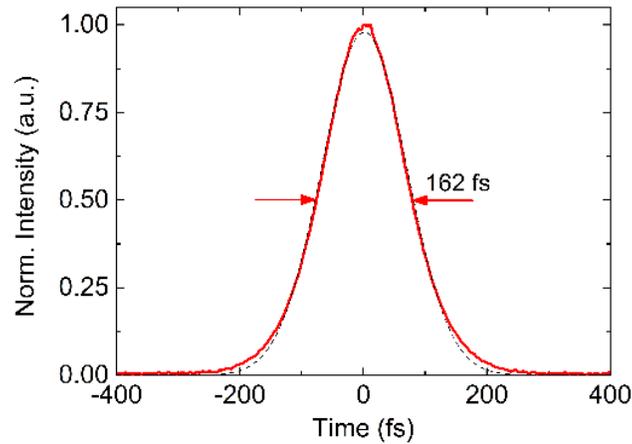

**Figure S4.** Measured intensity autocorrelation (red) and the Gaussian fit (dotted black) of the pulse.

**Distribution of MoS$_2$ crystals on coated ECF**

MoS$_2$ crystal distribution on a fiber section was determined by imaging the PL emission from the ECF sideways with a resolution of approximately 1 µm using a 10x microscope objective (Figure S5a). As described in the experimental setup, the objective and camera were translated along the fiber, and following the light propagation direction allowed us to obtain a set of images for a major part of the fiber. Some parts of the fiber were covered by the fixing mechanism and some mechanical space required by the incoupling and outcoupling objectives, where are not accessible.

A compound PL image was created after the mapping process and revealed in Figure S5b. The final image was split into 10 subsections for ease of display. The noise level of the camera was set and each bright spot with values above this threshold was considered as an active monolayer crystal. After analyzing the data, the total light emitted by each crystal together with their length were determined.

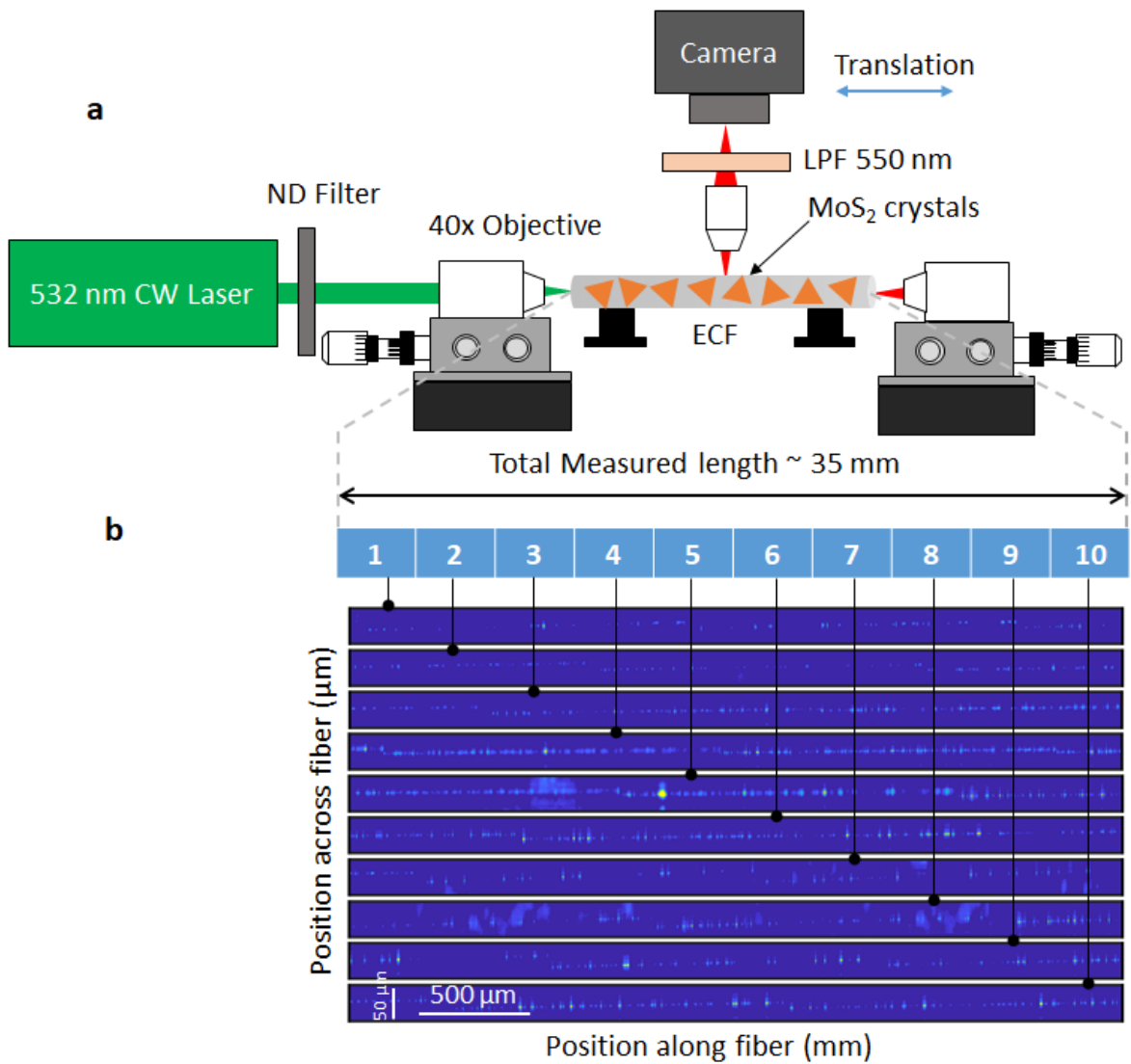

**Figure S5. Sideways PL measurement of MoS$_2$-crystal distribution. a**, Translational setup together with a sketch of the measured fiber sections which are numbered from 1 to 10. **b**, Compound PL side view image of the accessible 35 mm section of MoS$_2$-coated optical fiber, when the core is excited with a 532 nm laser, showing the distribution of MoS$_2$ crystals being excited by the fiber mode. The complete image was composed of sub-images, which are presented over ten lines for ease of presentation. Analysis of the images reveals a total length of 15.1 mm for coverage of 43.4 %.